\documentclass[a4paper,11pt]{article}
\usepackage[T1]{fontenc}
\usepackage{a4wide}
\usepackage{graphics}
\usepackage{epsfig}
\usepackage{wrapfig}
\usepackage{amssymb}
\usepackage{lscape}


\begin{document}

\title{Simulation of a fine grained GEM used in the PixiE Experiment
  \footnote{http://glastserver.pi.infn.it/pixie/pixie.html} \\
{\it PixiE Internal Report}}
\author{M. Del Prete \\
{\it\small INFN and University, Pisa} }
\date{15 May 2005}
\maketitle

\section*{Introduction}
\begin{figure}[ht]
 \begin{center}
  \mbox{\epsfig{file=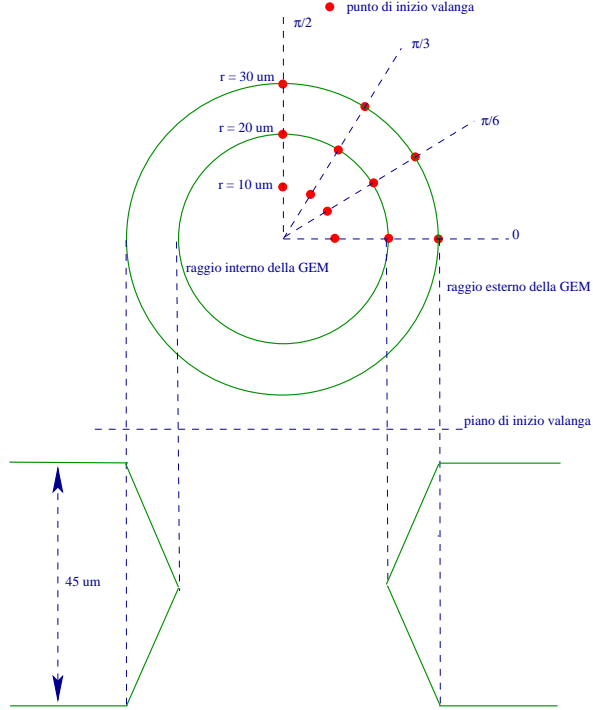,width=0.5\linewidth,angle=0}}
\caption{\small A section of the hole of GEM. The red circles indicate the
  starting points of electrons.}
\label{fig:fori}
\end{center}
\end{figure}

We have simulated the performances of a GEM with a large density of
multiplication holes. 
The elementary cell is an equilateral triangle whose side is 90$\mu m$. We
shall assume that this pattern extends in the (x,y) plane. 
At each vertex of the equilateral triangle there is a GEM hole 
with an external radius of 30$\mu
m$ and an internal  radius of 20$\mu m$.   
The reference frame used in this study has the origin of axis in the center of
the GEM hole with the z axis pointing to
the drift plane.
The geometry of the GEM is shown in figure \ref{fig:fori}.

In this short note we will describe the simulation of the GEM and, in
particular we will study the gain and diffusion of the charge for different
gas mixtures. 
This study has been performed to finalize the design of the PixiE Imager
Detector. 

We have started the simulation by generating single electrons in different
positions in the (x,y) plane at fixed z-coordinate. This is
the most elementary element through which we can simulate tracks and the
imaging performance of the detector. 
The process has been followed through multiplication in the large fields of the
GEM and diffusion of produced electrons reaching the readout
plane. 
Due to the cylindrical symmetry of the GEM the first quadrant ($x>0,y>0$, see
figure \ref{fig:fori}) has been selected to produce the 
coordinate ($r$,$\phi$) of the starting electrons at the quota of $40\;\mu
m$ (approximately $15\;\mu m$ over the top GEM): 
\[ r=10\cdot{n}\;\mu m\quad (n\;=1,\;2,\;3)\qquad 
\phi=\frac{\pi}{2}\cdot{\frac{(k-1)}{3}}\quad (k\;=\;1,\;2,\;3,\;4) \]
Where $\phi$ is the azimuthal angle. 

At each point we have generated 25 events. 
The study has been performed for the following gas mixtures: 
\begin{itemize}
\item 100\%$CO_2$ atm 0.5 ed 1 atm.
\item 20\%Ar/80\%DME, 50\%Ar/50\%DME, 80\%Ar/20\%DME ad 1 atm.
%\item 20\%Ne/80\%DME, 50\%Ne/50\%DME, 80\%Ne/20\%DME ad 1 atm.
\end{itemize}

\section*{Gain Study}
We have defined as absolute gain the number of electrons which reach the
quota $z\;=\;-40 \mu m$ (below the plane the GEM, approximately $15 \mu m$
below the bottom GEM plane). 
At this quota most of the multiplication processes at the GEM hole are done. 

However not all these electrons drift to the readout plane, 
some recombine and many
stick to the lower GEM plane (re-attachment). 
For this reason we have defined also an effective gain as the number
of electrons
which arrive at the quota $z\;=\;-200 \mu m$. The electrons reaching this
quota are considered to be collected by the read out plane. 

It's customary to describe the gain with a Polya:
\[
P_{n} =\frac{1}{b \cdot \overline{n}} \cdot 
\frac{1}{\Gamma\left(\frac{1}{b}\right)}\cdot
\left(\frac{n}{b \cdot 
\overline{n}}\right)^{\left(1/b \right) - 1}\cdot e^{-n/b \cdot \overline{n}}
\]

Where $b$ is an adjustable parameter and $\overline{n}$ is the average gain. 
We have used this formula and fit the data with the function: 
\[
F(n;p_{0},p_{1},p_{2}) = p_{2} \cdot n^{p_{0}-1} \cdot e^{-n/p_{1}}
\]
where $p_{2}$ is a normalization factor and the gain is the Polya mean, $G\;=\;p_{0}\;\cdot\;p_{1}$.

\begin{figure}[ht]
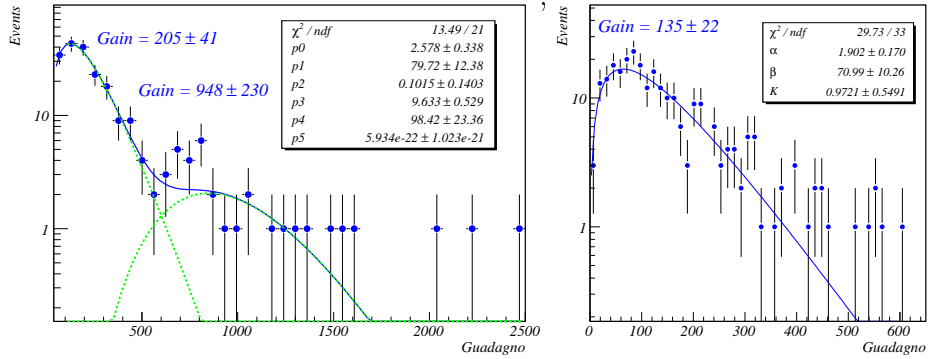

\begin{center}
  \mbox{\epsfig{file=figure/ga_90_480_4k_ARDME20-80_1atm.epsi,
                   width=0.3\linewidth,angle=270},
        \epsfig{file=figure/ge_90_480_4k_ARDME20-80_1atm.epsi,
                   width=0.3\linewidth,angle=270}}
\caption{\small Distribution of absolute (left) 
and effective (right) gain, the gas mixture
  is 50\% Argon \%50 DME. The GEM is operated of $480V$ with a collection field
  of $4KV/cm$.}
\label{fig:garco2}
\end{center}
\end{figure}

The gain distribution is different for different mixtures of gases. In
particular the absolute gain distribution is often wide with long tails 
and a description with a single Polya is not always satisfactory. Hence, we
have described the distribution with the sum of two Polya, of which 
the first one fits most of events and the second one accounts 
for the long tails. An example is shown in figure \ref{fig:garco2}. 
We have taken as the
average gain of the GEM the mean of the first Polya. 

\begin{figure}[h]
\begin{center}
  \mbox{\epsfig{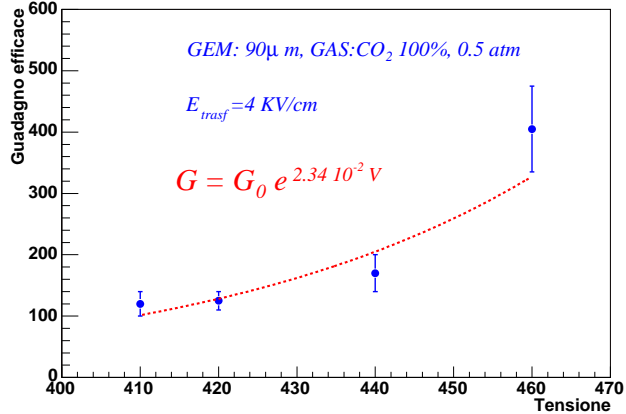}}
%{\resizebox*{8 cm}{!}
% {\includegraphics[angle=270]{figure/GvxHCO2100_0.5atm.epsi}}}
\caption{\small Dependence of effective gain on the GEM voltage. The simulation
  concerns a gas of 100\% $CO_2$ and a collection field of $4KV/cm$.}
\label{fig:geco2}
\end{center}
\end{figure}

\begin{figure}[h]
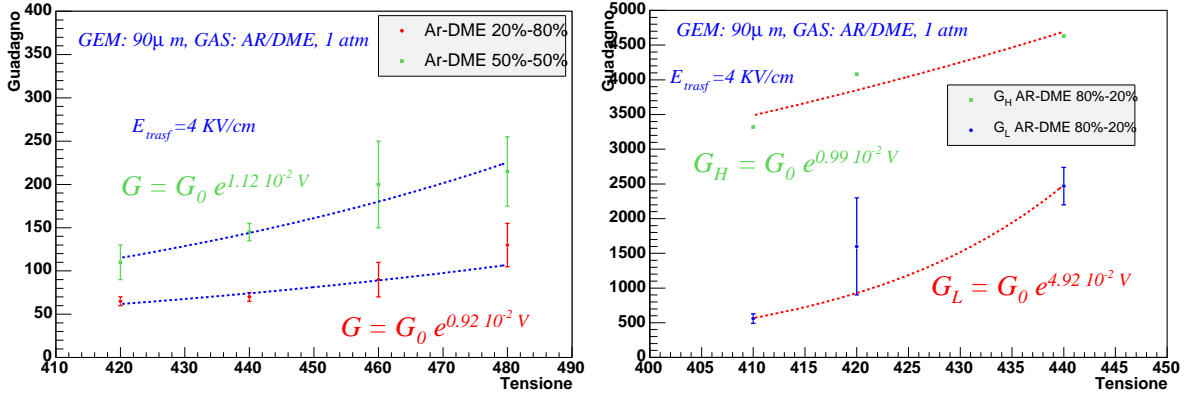

\begin{center}
  \mbox{\epsfig{file=figure/GvxHARDMEa_1atm.epsi,
                 width=0.33\linewidth,angle=270},
\epsfig{file=figure/GvxHARDME80-20_1atm.epsi,
                 width=0.33\linewidth,angle=270}
}
\caption{\small Left: gas mixtures of 20\%Ar-80\%DME and 50\%Ar-50\%DME. 
The plots show the dependence of the effective gain on the GEM voltage. 
The collection field is $4KV/cm$.
Right: gas mixture 80\%Ar-20\%DME with a large production of secondary 
avalanches. The figure shows the mean of effective gain as a function of the
GEM voltage and for a collection field of $4KV/cm$. 
The values of gain are the mean of first (gain low) and second 
(gain hight) Polya which fit the two peaks observed in the 
gain distribution.}
\label{fig:geardme1}
\end{center}
\end{figure}

Sometimes the distribution shows two clear maxima and a two Polya fit is
satisfactory. In this case the mean of the two Polya is the average gain of
the GEM under
analysis. Results are shown in the figure \ref{fig:geco2}, \ref{fig:geardme1}
for a collecting field $E_t\;=\;4KV/cm$ and two gas mixtures and in the table \ref{tab:co2051} for a collecting field of $E_t\;=\;5KV/cm$.

\vspace{0.5 cm}

\begin{table}[h]
\begin{center}
\begin{tabular}{c || c | c }
%%%%%100
$ \Delta V_{GEM} = 560\;V\quad E_{Trasf} = 5KV/cm$ &
 $p=0.5 atm$ & $p=1 atm $  \\
\hline \hline
$Gain_{eff}$ & 830 $\pm$ 100 & 70 $\pm$ 10   \\
\hline
$Gain_{ass}$ & 7420 $\pm$ 400 & 2400 $\pm$ 900  \\
\end{tabular}
\end{center}
\caption{\small  Average of effective and absolute gain of 100\% $CO_2$ at
  0.5 and 1 atm and for $\Delta V_{GEM}$ of $560V$ and collection field of $5KV/cm$.}
\label{tab:co2051}
\end {table}

The GEM gain increases with the voltage different GEM according to an exponential curve. 

\section*{Diffusion Study}

The study of the diffusion of the charge in the 
collecting region of the detector is important for two different issues. 

\begin{wrapfigure}[16]{r}{8 cm}
\begin{center}
\mbox{
\epsfig{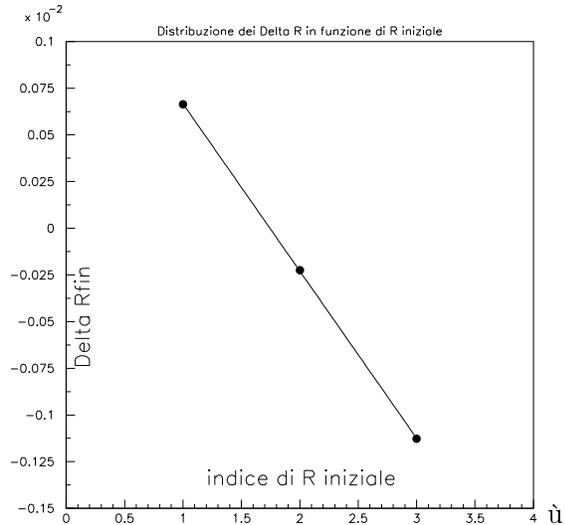}}
ù\caption{\small $\Delta R$ in function of $R_{in}$.}
\label{fig:diff8}
\end{center}
\end{wrapfigure}

Firstly to establish if the GEM keeps memory of the starting point of
the electrons, both in azimuth and radius (with respect 
to the center of the hole
where the avalanche occurs) with a better resolution than the granularity of
GEM's hole. 

For this, we have studied the position of the barycentre of charge 
arrived on readout plane
(barycentre of the avalanche) as a
function of the position of the starting point. 
 
The second point is the RMS of Gaussian 
distribution of charge in the collection gap which is related with the spatial
resolution of the detector.

The average position of the collected charge indicates where the multiplication
occurs at the GEM hole. 

We have considered 
\[
\Delta R\;=\;R_{fin}\;-\;R_{in}
\]
where $R_{in}\;=\;r$ and $R_{fin}$ is the radius of the average charge at the
quota $z\;=\;-200 \mu m$. 
First of all we have studied the dependence of $\Delta R$ on $R_{in}$

\begin{figure}[ht]
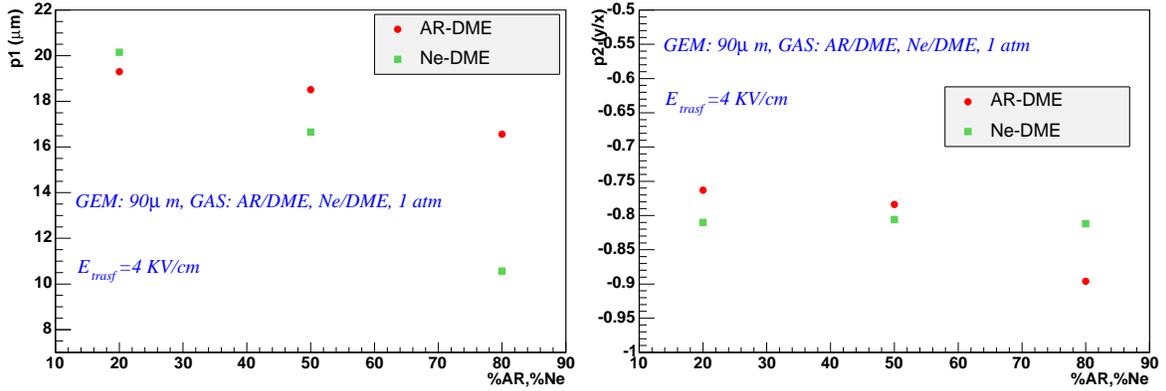

\begin{center}
  \mbox{\epsfig{file=figure/AR-Ne_p1.epsi,
                 width=0.33\linewidth,angle=270},
\epsfig{file=figure/AR-Ne_p2.epsi,
                 width=0.33\linewidth,angle=270}}
\caption{\small The values of $p1$ ($\mu m$) and $p2$ in function of percentage 
of Argon and Neon.}
\label{fig:p1}
\end{center}
\end{figure}

The figure \ref{fig:diff8} shows an example of such a dependence.
$ \Delta R$ is a linear function of $ R_{in}$ and a parameterization: 
$ \Delta R\;=P_2 R_{in}\;+\;P_1 $ 
with $P_2 \approx -1$ and $P_1 \approx 20
\mu m$ is a good fit for all simulations (figure \ref{fig:p1}). 
For a ideal GEM $\Delta R = 0$ in fact this means that the position (in radius)
of the multiplied charge is the same as the initial electron (the GEM does
not disturb the image). The results of the fit shows instead that the average
collected charge position is independent of the starting position ($P_2
\approx -1$) and that the multiplication occurs at the radius $r \approx 20-30
\mu m$ (the lower external radius). 

The last step is the study of the dispersion of the avalanche's charge after
it drifted to the collection electrodes. 

We have averaged the RMS of the events produced at each point and verified
that its value is independent of the position of primary electron. Hence we
have averaged the RMS of all events at all points, to improve the statistics
and studied its dependence on the GEM voltage. 
Since, again, we have found no dependence, we have average on all events for a
defined gas composition. 

The results for 100\%$CO_{2}$ gas at $0.5\;atm$ are:
\begin{itemize}
\item $RMS_{p}\;=\;12.5\;\mu m$ 
\item  $RMS_{sp}\;=\;17.67\;\mu m$ 
\end{itemize}

\begin{figure}[ht]
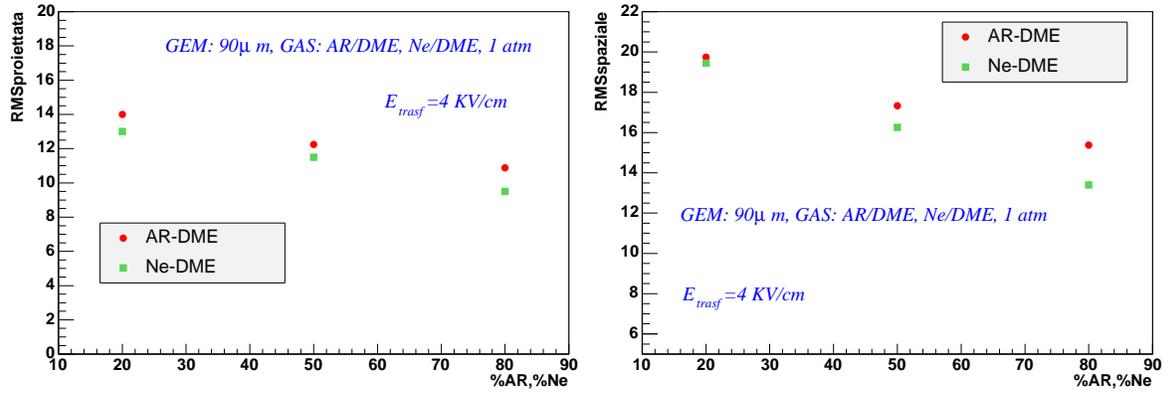

\begin{center}
  \mbox{\epsfig{file=figure/AR-Ne_RMSp.epsi,
                 width=0.33\linewidth,angle=270},
\epsfig{file=figure/AR-Ne_RMSsp.epsi,
                 width=0.33\linewidth,angle=270}}
\caption{\small Left we show the projected $RMS$ ($\mu m$) averaged on all events for
  two gas mixtures while on the right the spatial $RMS$ ($\mu m$).}
\label{fig:AR-Ne_RMSp}
\end{center}
\end{figure}

The results for Argon and Neon mixtures are shown in figure
\ref{fig:AR-Ne_RMSp}, the RMS decreases mildly with increasing percentage of
Argon and Neon. 
  
\section*{Acknowledgments} 
I would like ti thank G.Spandre for the help and continuous advice on my work
and R.Veenhof for his support in the use of the simulation program Garfield
and also for many advice on how to perform reliable simulations.
\end{document}